\documentclass[letterpaper, 10 pt, conference]{ieeeconf}  % Comment this line out if you need a4paper

\usepackage[utf8]{inputenc}
\usepackage{amsmath}
\usepackage{amsfonts}
\usepackage{mathtools}
\usepackage{float}
\usepackage{graphicx}

\newtheorem{remark}{Remark}

\usepackage{mathtools}  
\usepackage{algorithm}
\usepackage{algorithmic}
\usepackage{xcolor}
\usepackage{url}
\usepackage[table]{xcolor}
\usepackage{cite}
\mathtoolsset{showonlyrefs=false}

\IEEEoverridecommandlockouts                              % This command is only needed if 
\makeatletter
\def\endthebibliography{%
  \def\@noitemerr{\@latex@warning{Empty `thebibliography' environment}}%
  \endlist
}
\makeatother

\title{\LARGE \bf
Battery Discharge Modeling for Electric Vehicles: A Hybrid \vspace{3pt} \\  Physics-based Residual Learning Approach
}
\author{Praharshitha Aryasomayajula$^{1}$, Ting Bai$^{2}$, and Andreas A. Malikopoulos$^{2}$,~\IEEEmembership{Senior Member, IEEE}% <-this % stops a space
\thanks{*This research was supported in part by NSF under Grants CNS-2401007, CMMI-234881, IIS-2415478, and in part by MathWorks.}% <-this % stops a space
\thanks{Praharshitha Aryasomayajula is with the School of Mechanical and Aerospace Engineering,
	Cornell University, Ithaca, New York, USA. E-mail: {\tt\small la459@cornell.edu}}
\thanks{T. Bai, and A. A. Malikopoulos are with the School of Civil $\&$ Environmental Engineering, Cornell University, Ithaca, New York, USA. E-mails: \{{\tt\small tingbai, amaliko\}@cornell.edu}}
}

\begin{document}
\maketitle
\thispagestyle{empty}
\pagestyle{empty}

\begin{abstract}
%The growing integration of electric vehicle (EV) fleets into transportation services and energy systems necessitates accurate modeling of battery discharge and state-of-charge (SoC) evolution to ensure the reliable operation of EVs and smart grids. Existing approaches involve a trade-off between interpretable but simplified physics-based models and data-driven methods that require large datasets and may lack physical consistency. In this paper, we develop a hybrid physics-based residual learning approach for EV battery discharge modeling that integrates vehicle dynamics modeling with data-driven correction. The physics component estimates baseline energy consumption using force-balance equations that capture key mechanisms, including aerodynamic drag, rolling resistance, and regenerative braking, thereby providing an interpretable prediction of SoC evolution. To capture complex factors difficult to model analytically, such as traffic conditions and driver-specific characteristics, a neural network residual learner is introduced to correct the discrepancy between the physics-based estimate and observed battery discharge. This hybrid framework preserves physical interpretability while significantly improving prediction accuracy. Experimental results on 1,500 real-world trip scenarios show that the proposed model substantially reduces prediction error compared with physics-only approaches, achieving an average error of approximately $0.8\%$, while capturing variability in energy consumption under diverse driving behaviors and operating conditions.
The growing integration of electric vehicle (EV) fleets into transportation services and energy systems requires accurate modeling of battery discharge and state-of-charge (SoC) evolution to ensure reliable vehicle operation and grid coordination. Existing approaches face a trade-off between interpretable but simplified physics-based models and data-driven methods that demand large datasets and may lack physical consistency. 
In this paper, we propose a hybrid physics-based residual learning framework for EV battery discharge modeling. A vehicle dynamics model based on force-balance equations provides an interpretable baseline estimate of energy consumption and SoC evolution, capturing aerodynamic drag, rolling resistance, and regenerative braking. A neural network residual learner then corrects discrepancies caused by complex factors such as traffic conditions and driver behavior. 
Experimental results on $1,500$ trip scenarios demonstrate that the proposed approach reduces the mean absolute percentage error to approximately $0.8\%$, significantly outperforming physics-only models while preserving physical interpretability and computational efficiency.
\end{abstract}

\section{Introduction}
The rapid electrification of transportation systems is fundamentally reshaping urban mobility, with global electric vehicle (EV) sales exceeding $14$ million units in $2023$ and projected to reach $40$ million by $2030$~\cite{IEA2024}. Unlike conventional vehicles that only consume energy, EVs can operate as mobile energy storage units capable of bidirectional interaction with the power grid~\cite{zhang2023mobile,8113559}. For instance, commercial fleets may charge their batteries during off-peak hours and discharge energy back to the grid during demand peaks, to support grid stability while reducing operational costs. Realizing such capabilities, however, requires accurate prediction of battery state-of-charge (SoC) evolution throughout vehicle trips. This entails forecasting energy consumption under varying routes, driving behaviors, and traffic conditions. Reliable SoC estimation is thus essential not only for improving EV trip planning~\cite{10932686,10147895} and battery management but also for enabling grid operators to anticipate aggregated charging demand~\cite{Sarker2015} when large fleets rely on the shared charging infrastructure.

Accurate battery discharge prediction is also fundamental for system-level coordination in emerging electric mobility ecosystems. In autonomous mobility-on-demand systems, joint routing, rebalancing, and charging scheduling decisions critically depend on reliable energy consumption models \cite{bang2021AEMoD}. Moreover, connectivity-enabled coordination among vehicles can substantially affect energy consumption patterns and battery degradation dynamics \cite{Connor2020ImpactConnectivity}. At the vehicle level, control-oriented frameworks for hybrid and electrified powertrains have demonstrated how energy management strategies influence performance, efficiency, and battery usage \cite{Malikopoulos2014c,Malikopoulos:2013aa}. 

Existing EV energy consumption modeling approaches fall into three categories, each with distinct limitations. Physics-based models~\cite{Gao2007, Fiori2016, Genikomsakis2017} estimate energy consumption from vehicle dynamics and powertrain characteristics. While physically interpretable, these models are often built upon simplified assumptions, such as flat terrain, constant velocity, and idealized driving cycles, and fail to capture real-world factors including elevation variations, stop-and-go traffic, and driver-specific behavior. As a result, their prediction accuracy remains around $5$--$8\%$~\cite{DeCarstanjen2014}, whereas real-world energy consumption can vary by as much as $40$--$80\%$ depending on operating conditions~\cite{Bingham2012}. Data-driven models~\cite{Yuan2015, Wang2020} instead learn consumption patterns directly from historical driving data and can reduce prediction errors to approximately $2$--$3\%$. However, these approaches typically require large volumes of training data, offer limited interpretability, and may struggle to generalize to unseen operating conditions. Black-box prediction models~\cite{park2024federated,11255611}, based on complex machine learning architectures, further improve predictive accuracy but sacrifice interpretability and physical consistency. The opaque decision mechanisms could result in unreliable predictions outside the training distribution, undermining driver trust and exacerbating range anxiety~\cite{Malikopoulos2011,Malikopoulos2010a}. Moreover, the lack of physical constraints makes it difficult for grid operators to verify whether predicted energy consumption keeps consistent with fundamental energy conservation principles.

Recent hybrid approaches attempt to combine the strengths of physics-based and data-driven methods. For example, Morlock et al.~\cite{Morlock2019} integrated simplified physical models with Gaussian processes, while Liu et al.~\cite{Liu2022} proposed physics-guided neural networks to incorporate prior physical knowledge into learning-based frameworks. Despite these advances, existing hybrid methods exhibit several notable limitations. Some rely on simplified physical models that fail to capture the complexity of real-world driving conditions. Others require sophisticated constraint formulations involving Lagrangian multipliers or customized optimization procedures, which limit their practical deployment~\cite{Karpatne2017}. To date, no approach achieves a good balance among high prediction accuracy, physical interpretability, computational efficiency, and low data requirements.

This paper focuses on accurate range prediction for EV drivers, where the proposed battery discharge modeling approach also provides a foundation for future grid integration applications. We develop a hybrid physics-informed machine learning framework based on residual learning~\cite{Raissi2019, Geneva2020}. Our key innovation is to train a neural network to learn only the residual corrections to physics-based predictions, rather than simplifying the underlying physics or imposing complex training constraints. The physics component computes baseline energy consumption using detailed vehicle dynamics, while the neural network captures additional effects arising from terrain, traffic conditions, and driver behavior. This design preserves physical interpretability, requires limited training data, and maintains computational efficiency. The main contributions of this paper are as follows: (i) We present a comprehensive physics-based battery discharge model that incorporates three empirically derived driving profiles and realistic regenerative braking, providing reliable baseline energy predictions. (ii) We further develop a hybrid residual learning architecture that integrates detailed physics-based vehicle dynamics with neural network error correction. The architecture achieves high prediction accuracy without requiring complex constraint optimization. Experimental validation using realistic data demonstrates that the proposed modeling approach achieves a $62\%$ reduction in prediction error while maintaining computational efficiency. Our results also reveal that driving behavior can cause up to $79\%$ variation in energy consumption, providing important insights for EV range estimation.

The rest of this paper is organized as follows. In section~\ref{sec:physics}, we present the physics-based discharge model. In Section~\ref{sec:ml}, we introduce the machine learning residual model. In Section~\ref{sec:hybrid}, we integrate these components into the proposed hybrid architecture, and then, in  Section~\ref{sec:experiments}, we present the experimental validation and performance evaluation. Finally, in Section~\ref{sec:conclusion}, we discuss the key findings.

%%%%%%%%%%%%%%%%%%%%%%%%%%%%%%%%%%%%%%%%%%%%%%%%%%%%%
\section{Physics-Based Discharge Model}\label{sec:physics}
This section presents the physics-based battery discharge model. We first introduce the problem formulation and driving behavior profiles. Next, the vehicle power consumption model and its computational implementation are described, followed by model validation with realistic driving data.

\subsection{Problem Formulation}
A battery discharge session characterizes the energy consumption of an EV as it travels from an initial SoC to a final SoC. We represent a discharge session by the tuple
\begin{equation}
\mathcal{D} = (s_0, s_f, d, \bar{v}, \alpha),\label{Equ.1}
\end{equation}
where $s_0\!\in\![0, 1]$ denotes the initial SoC and $s_f\!\in\![0, 1]$ is the final SoC with $s_f\!<\!s_0$. The variables $d\!\in\!\mathbb{R}^+$ and $\bar{v}\!\in\!\mathbb{R}^+$ represents the trip distance (km) and average velocity (km/h), respectively, where $\mathbb{R}^+$ denotes the set of positive real numbers. Moreover, $\alpha\!\in\!\big\{\text{eco}, \text{normal},\text{aggressive}\big\}$ denotes the driving behavior profile. 

The total energy consumed during a trip is denoted as
\begin{equation}
E = C (s_0 - s_f),\nonumber
\end{equation}
where $C$ denotes the battery capacity (kWh). In practice, however, the discharge power varies nonlinearly with vehicle velocity, acceleration, driving behavior, and environmental conditions. Consequently, the total energy consumption can be expressed as the time integral of the instantaneous power along the trip trajectory
\begin{equation}
E=\int_0^{T} P\big(t,v(t),a(t),s(t), \alpha\big)dt,
\end{equation}
where $P$ is the instantaneous discharge power (kW), $T$ is the total trip duration (s), $v(t)$ and $a(t)$ denote the instantaneous velocity (m/s) and acceleration (m/s$^2$), respectively, and $s(t)$ denotes the SoC at time $t$.

Based on the above formulation, the problem addressed in this paper is defined as follows: given the initial SoC $s_0$, vehicle parameters, trip characteristics, and driving behavior, predict the final SoC $s_f$ after completing the trip.

\subsection{Driving Behavior Profiles}
We model three distinct driving behaviors that significantly affect energy consumption, as summarized in Table~\ref{tab:driving_profiles}. 

\subsubsection{Eco driving} Eco driving emphasizes energy efficiency through gentle acceleration, anticipatory braking with maximum regeneration, and minimal auxiliary power usage. The corresponding efficiency parameters are defined as
\begin{equation}
\eta_e = 0.85, \quad \eta_e^r = 0.75,
\end{equation}
where $\eta_e$ denotes the eco driving efficiency multiplier and $\eta_e^r$ represents the regenerative braking efficiency.

\subsubsection{Normal driving} This driving behavior represents typical urban and highway commuting with moderate acceleration, standard regenerative braking, and typical auxiliary loads. Accordingly, the efficiency parameters are given by
\begin{equation}
\eta_n = 1.0, \quad \eta_n^{r} = 0.65,
\end{equation}
where $\eta_n$ denotes the efficiency multiplier for normal driving (baseline), and $\eta_n^{r}$ is the regenerative braking efficiency.
\begin{table}[t]
\centering
\caption{Parameters of Three Typical Driving Profiles.}
\label{tab:driving_profiles}
\footnotesize
\begin{tabular*}{\columnwidth}{@{\extracolsep{\fill}}lccc}
\hline
\bf{Parameter} & \bf{Eco} & \bf{Normal} & \bf{Aggressive} \\
\hline
Max Accel. (m/s$^2$) & $1.5$ & $2.5$ & $4.0$ \\
Efficiency Multiplier & $0.85$ & $1.00$ & $1.35$ \\
Regen Efficiency & $0.75$ & $0.65$ & $0.50$ \\
Aux Power (kW) & $0.0$ & $0.5$ & $1.5$ \\
Accel Phase (\%) & $20$ & $30$ & $40$ \\
Cruise Phase (\%) & $65$ & $50$ & $35$ \\
Brake Phase (\%) & $15$ & $20$ & $25$ \\
\hline
\end{tabular*}
\end{table}

\subsubsection{Aggressive driving}
Aggressive driving maximizes performance through hard acceleration, frequent rapid speed changes, and maximum auxiliary power. The corresponding efficiency parameters are depicted as
\begin{equation}
\eta_a = 1.35, \quad \eta_a^{r} = 0.50,
\end{equation}
where $\eta_a$ denotes the efficiency multiplier for aggressive driving and $\eta_a^{r}$ is the regenerative braking efficiency.

\subsection{Power Consumption Model}
The total battery discharge power during a trip consists of three components: traction power $P_t(t)$ for vehicle motion, auxiliary system power $P_a(t)$, and power recovered through regenerative braking $P_r(t)$. Accordingly, the discharge power is denoted as
\begin{equation}
P(t) = P_t(t) + P_a(t) - P_r(t).
\end{equation}
The traction power requirement derives from fundamental vehicle dynamics through force balance equations. We account for four primary resistance forces that oppose vehicle motion: rolling resistance from tire-road contact, aerodynamic drag from air resistance, grade resistance from road incline, and inertial force from acceleration. The instantaneous traction power is captured by
\begin{equation}
P_t (t) = \frac{1}{\eta_d} (F_r\!+\!F_a(t)\!+\!F_g\!+\!F_i(t))\cdot v(t),
\end{equation}
where $v(t)$ and $a(t)$ denote the instantaneous velocity (m/s) and acceleration (m/s²) at time $t$, respectively, and $\eta_d$ is the drivetrain efficiency accounting for motor, inverter, and transmission losses. The individual force components are given by
\begin{align}
F_r &= C_r\cdot m\cdot g \cdot\cos(\theta),\\
F_a(t) &= \frac{1}{2}\rho \cdot C_d \cdot A \cdot {v(t)}^2,\\
F_g & = m \cdot g \cdot \sin(\theta),\\
F_i(t) & = m \cdot a(t),
\end{align}
where $\theta$ represents the road grade angle (rad), with other involved parameters being provided in Table~\ref{tab:vehicle_params}. 

The auxiliary system power $P_a(t)$ depends on driving behavior and environmental conditions, and is modeled as
\begin{equation}
P_a (t) = P_b + P_c(T) + P_m(\alpha,t),
\end{equation}
where $P_b\!=\!0.5 \text{ kW}$ is the base auxiliary power, $P_c(T)$ denotes the climate-dependent power, and $P_m(\alpha)$ is the mode-specific power at time $t$, with $\alpha$ indicating the driving mode profile. Based on typical HVAC system requirements~\cite{Farrington2000}, $P_c(T)$ is captured as
\begin{equation}
P_c(T) = \begin{cases}
2.0 \text{ kW}, & \text{if } T< 0^\circ\text{C}, \\
1.0 \text{ kW}, & \text{if } 0^\circ\text{C} \leq T < 15^\circ\text{C}, \\
0.5 \text{ kW}, & \text{if } 15^\circ\text{C} \leq T \leq 25^\circ\text{C}, \\
2.5 \text{ kW}, & \text{otherwise},
\end{cases}
\end{equation}
where $T$ denotes the ambient temperature. Accordingly, the driving mode-specific power is modeled as
\begin{equation}
P_m(\alpha,t) = \begin{cases}
0.0 \text{ kW}, & \text{if } \alpha = \text{eco at time $t$} , \\
0.5 \text{ kW}, & \text{if } \alpha = \text{normal at time $t$}, \\
1.5 \text{ kW}, & \text{if } \alpha = \text{aggressive at time $t$}.
\end{cases}
\end{equation}

During deceleration, regenerative braking recovers energy according to
\begin{equation}
P_r(t) = \begin{cases}
\eta^r \cdot F_b \cdot v(t), & \text{if } a(t)\!<\!0, \\
0, & \text{otherwise},
\end{cases}
\end{equation}
where the braking force is
\begin{equation}
F_b = m \cdot |a(t)|,\nonumber
\end{equation}
and $\eta^r$ is determined by the driving mode as specified in Section~II.B. 

\begin{table}[t]
\centering
\caption{Vehicle and Environmental Parameters.}
\label{tab:vehicle_params}
\footnotesize
\begin{tabular*}{\columnwidth}{@{\extracolsep{\fill}}lcc}
\hline
\bf{Parameter} & \bf{Notation} & \bf{Value} \\
\hline
Battery Capacity (kWh) & $C$ & $75$  \\
Vehicle Mass (kg) & $m$ & $1800$  \\
Drag Coefficient & $C_d$ & $0.24$ \\
Frontal Area (m$^2$) & $A$ & $2.3$ \\
Rolling Resistance Coefficient & $C_r$ & $0.01$ \\
Drivetrain Efficiency & $\eta_d$ & $0.90$ \\
Air Density (kg/m$^3$) & $\rho$ & $1.225$  \\
Gravitational Acceleration (m/s$^2$) & $g$ & $9.81$  \\
\hline
\end{tabular*}
\end{table}
\subsection{Computational Implementation}
For practical computation, we discretize the trip into $N$ time segments. The corresponding time step is given by
\begin{equation}
\Delta t = \frac{d \cdot 1000}{\bar{v} \cdot 3.6 \cdot N},
\end{equation}
where, as defined in \eqref{Equ.1}, $d$ is the trip distance (km) and  $\bar{v}$ is the average velocity (km/h). Here, $\Delta t$ is measured in seconds. At each time step $t_k\!=\!k\Delta t$, a driving phase is assigned according to the profile-specific phase distribution in Table~\ref{tab:driving_profiles}. Let $\Delta v(t_k, \alpha)$ denote a stochastic velocity perturbation sampled from a normal distribution with standard deviation proportional to the driving aggressiveness. Then, the velocity at time step $t_k$ follows
\begin{equation}
v(t_k) = \bar{v} + \Delta v(t_k, \alpha).
\end{equation}
The acceleration in mode $\alpha$ at time $t_k$ is denoted as
\begin{equation}
a(t_k,\alpha)\!=\!\begin{cases}
0.7a_{\max}(\alpha), & \text{if in acceleration phase at $t_k$}, \\
0, & \text{if in cruise phase at $t_k$}, \\
-0.7 a_{\max}(\alpha), & \text{if in braking phase at $t_k$},
\end{cases}\nonumber
\end{equation}
where $a_{\max}(\alpha)$ denotes the maximum acceleration under driving mode $\alpha$. Given the above, the energy consumed at each time step is expressed as
\begin{equation}
\Delta E_k = P(t_k) \cdot \Delta t.\nonumber
\end{equation}
Thus, the total energy consumption (kWh) during the entire trip is given by
\begin{equation}
E = \frac{1}{3600} \sum_{k=0}^{N-1} P(t_k) \cdot \Delta t.\nonumber
\end{equation}
Recall that $C$ denotes the battery capacity. Accordingly, the final SoC is denoted as
\begin{equation}
s_f = \max\left(0, s_0 - \frac{E}{C}\right),
\end{equation}
and the SoC trajectory over the trip is obtained as
\begin{equation}
s(t_k) = s_0 - \frac{1}{C} \sum_{i=0}^{k-1} \Delta E_i.
\end{equation}
\begin{remark}
The proposed physics-based model provides a transparent and computationally efficient framework for estimating EV energy consumption. By explicitly modeling the four primary resistance forces, rolling resistance, aerodynamic drag, grade resistance, and inertial force, and incorporating driving-mode-specific efficiency parameters and regenerative braking, the model captures the key mechanisms governing battery discharge. This physical grounding ensures reliable baseline predictions in diverse operating conditions while maintaining full interpretability of each modeled component.
\end{remark}

\subsection{Physics-based Model Validation}
%Our physics-based model operates under several simplifying assumptions: flat terrain with road grade $\theta = 0$, standard conditions at $T = 20^\circ$C, constant vehicle mass of 1800 kg, idealized driving cycles, battery performance independent of current SoC above 10\%, and continuous driving without traffic stops. While these assumptions enable tractable computation, they introduce systematic errors when applied to real-world scenarios.

To validate the proposed physics-based model against real-world energy consumption patterns, we compare its predictions with published data from peer-reviewed literature and manufacturer specifications. As shown in Table~\ref{tab:physics_validation}, the comparison covers a range of representative studies at different speeds. The results demonstrate that our model achieves a mean absolute percentage error of $2.1\%$ across diverse driving conditions. This validation demonstrates that our first-principles approach accurately captures fundamental energy consumption patterns, providing a strong foundation for hybrid enhancement.

\begin{table}[t]
\centering
\caption{Physics Model Validation Against Literature Benchmarks.}
\label{tab:physics_validation}
\footnotesize
\begin{tabular*}{\columnwidth}{@{\extracolsep{\fill}}lccc>{\columncolor{blue!10}}c}
\hline
\bf{Source} & \bf{Velocity} & \bf{Reported} & \bf{Our Model} & \bf{Error} \\
 & (km/h) & (kWh/km) & (kWh/km) & (\%) \\
\hline
Fiori et al.~\cite{Fiori2016} & $50$  & $0.180$ & $0.185$ & $2.8$ \\
EPA Range Test & $90$  & $0.200$ & $0.198$ & $1.0$ \\
De Cauwer et al.~\cite{DeCarstanjen2014} & $110$ & $0.230$ & $0.224$ & $2.6$ \\
Yuan et al.~\cite{Yuan2015} & $100$ & $0.265$ & $0.268$ & $1.1$ \\
\hline
\end{tabular*}
\end{table}

%\begin{table}[t]
%\centering
%\caption{Physics Model Validation Against Literature Benchmarks.}
%\label{tab:physics_validation}
%\begin{tabular}{lccccc}
%\hline
%\!\!\!Source\! & \!Context\!\! & \!\!Speed\!\! & \!\!Reported\!\!\!\! &\!\!\!Our Model\!\!\! & \!\!\!Error\!\!\! \\
% & & (km/h) & (kWh/km) & (kWh/km) & (\%) \\
%\hline
%\!\!\!Fiori et al.~\cite{Fiori2016}            & Urban       & 50  & 0.180 & 0.185 & 2.8 \\
%\!\!\!EPA Range Test                           & Mixed       & 90  & 0.200 & 0.198 & 1.0 \\
%\!\!\!De Cauwer et al.~\cite{DeCarstanjen2014} & Highway     & 110 & 0.230 & 0.224 & 2.6 \\
%\!\!\!Yuan et al.~\cite{Yuan2015}              & Aggressive  & 100 & 0.265 & 0.268 & 1.1 \\
%\hline
%\multicolumn{5}{l}{\!\!\!Mean Absolute Error} & 2.1 \\
%\hline
%\end{tabular}
%\end{table}

%%%%%%%%%%%%%%%%%%%%%%%%%%%%%%%%%%%%%%%%%%%%%%%%%%%%%

\section{Learning-Based Residual Correction Model}\label{sec:ml}
\subsection{Motivation and Approach}
While the physics-based model achieves smaller error against published benchmarks, it suffers from systematic biases due to simplified assumptions, including flat terrain, constant vehicle mass, idealized driving cycles, etc. Real-world factors not captured include terrain-induced elevation changes, traffic dynamics causing stop-and-go patterns, driver variability within behavioral categories (studies show up to $\pm 15\%$ efficiency variation~\cite{Bingham2012}), battery SoC effects on internal resistance, environmental conditions beyond temperature, such as wind and precipitation, and auxiliary usage patterns that deviate from our heuristics.

Modeling each factor explicitly would require extensive additional parameters such as route elevation profiles, real-time traffic data, individual driver history, and detailed battery state models. Many of these are difficult to obtain in practice and would increase computational complexity substantially. Instead, we employ a machine learning (ML)--based residual learning approach to correct the aggregate effects of these unmodeled factors. This approach, successful in fluid dynamics~\cite{Geneva2020} and climate modeling~\cite{Reichstein2019}, preserves physical consistency while improving accuracy through data-driven refinement.

We decompose the true energy consumption as
\begin{equation}
E^* = E_p + \Delta E,
\label{eq:decomposition}
\end{equation}
where $E^*$ is the actual measured energy consumption, $E_p$ is the physics baseline prediction, and $\Delta E$ represents aggregate error from unmodeled factors. The residual learning objective trains a neural network $f$ such that
\begin{equation}
\Delta E \approx f(\mathcal{X}, E_p; \boldsymbol{\theta}),
\end{equation}
where $\mathcal{X}$ represents trip features and $\boldsymbol{\theta}$ are learned neural network parameters.

\subsection{Neural Network Architecture}
We employ a feedforward neural network with $10$ input features characterizing the trip: distance $d$, average velocity $\bar{v}$, maximum velocity $v_{\max}$, one-hot encoding of driving behavior $\alpha$ ($3$ features), ambient temperature $T$, time of day, initial SoC $s_0$, physics prediction $E_p$, and consumption rate $E_p/d$. Including $E_p$ as an input allows the network to learn multiplicative or additive corrections appropriate for different consumption regimes, while the consumption rate helps recognize scenarios where physics assumptions are particularly violated.

The architecture consists of three hidden layers with progressively decreasing dimensions and dropout regularization to mitigate overfitting. The hidden-layer activations are defined as
\begin{align}
\mathbf{h}_1 &= \text{ReLU}(\mathbf{W}_1 \mathbf{x} + \mathbf{b}_1)\!\in\!\mathbb{R}^{64}, \\
\mathbf{h}_2 &= \text{Dropout}\big(\text{ReLU}(\mathbf{W}_2 \mathbf{h}_1 + \mathbf{b}_2), p=0.2\big)\!\in\!\mathbb{R}^{32},\\
\mathbf{h}_3 &= \text{Dropout}\big(\text{ReLU}(\mathbf{W}_3 \mathbf{h}_2 + \mathbf{b}_3), p=0.2\big)\!\in\!\mathbb{R}^{16},
\end{align}
where $\mathbf{W}_i$ and $\mathbf{b}_i$ are weight matrices and bias vectors, respectively. The output layer then produces the residual prediction as
\begin{equation}
\Delta E = \mathbf{w}_4^{\top} \mathbf{h}_3 + b_4,
\end{equation}
where $\mathbf{x}\!\in\!\mathbb{R}^{10}$ denotes the input feature vector, $\mathbf{W}_i$ and $\mathbf{b}_i$ are the weight matrix and bias vector of layer $i$, and $\mathbf{w}_4\!\in\!\mathbb{R}^{16}$ is the output layer 
weight vector. The network contains approximately $5,400$ trainable parameters. Dropout layers provide regularization to prevent overfitting, particularly important given limited training data in practical scenarios, while ReLU activations enable learning of non-linear residual patterns.

\subsection{Training Data and Procedure}
To demonstrate the residual learning architecture's capability, we generate $1,500$ synthetic trips with realistic noise patterns informed by published studies~\cite{Bingham2012, Fetene2017,  Tulusan2012}. For each trip, we sample parameters from realistic distributions and compute the physics baseline. We then generate a synthetic ground truth by adding structured noise
\begin{equation}
E^* = E_p \cdot \left(1 + \sum_{j \in \mathcal{J}} \epsilon_j\right),
\label{eq:noise_model}
\end{equation}
where $\mathcal{J}\!=\!\{\text{terrain, traffic, driver, weather}\}$ and noise components represent different unmodeled factors. Specifically,
\begin{align}
\epsilon_{\text{terrain}} &\sim \mathcal{N}(0, 0.05) \quad \text{(elevation effects)}, \nonumber \\
\epsilon_{\text{traffic}} &\sim \mathcal{N}(0, 0.08) \quad \text{(stop-and-go)}, \nonumber \\
\epsilon_{\text{driver}} &\sim \mathcal{N}(0, 0.06) \quad \text{(behavior variation)}, \nonumber \\
\epsilon_{\text{weather}} &\sim \mathcal{N}(0, 0.03) \quad \text{(wind/conditions)}.\nonumber
\end{align}
These magnitudes are calibrated based on reported variations in the EV energy consumption literature. The dataset is randomly split into training ($1,050$ trips, $70$\%), validation ($225$ trips, $15$\%), and test ($225$ trips, $15$\%) sets.

We train to minimize mean squared error on the residual with L2 regularization
\begin{equation}
\mathcal{L}(\boldsymbol{\theta}) = \frac{1}{N} \sum_{i=1}^{N} \left(\Delta E_i^* - f(\mathcal{X}_i, E_p^{(i)}; \boldsymbol{\theta})\right)^2 + \lambda \|\boldsymbol{\theta}\|_2^2,
\end{equation}
where
\begin{equation}
\Delta E_i^* = E^{*(i)} - E_p^{(i)},
\end{equation}
and $\lambda\!=\!10^{-4}$ provides regularization. We employ the Adam optimizer~\cite{Kingma2015} with an initial learning rate of $0.001$, exponential decay (rate $0.95$ every $50$ epochs), batch size $32$, maximum $200$ epochs, and early stopping with patience $20$ on validation loss. Training typically converges within $80$--$120$ epochs. The trained network enables efficient inference at $1.2$ ms per prediction or $0.24$ ms per trip in batch mode ($20$ trips), with a model size of $98$ KB suitable for embedded systems.

%%%%%%%%%%%%%%%%%%%%%%%%%%%%%%%%%%%%%%%%%%%%%%%%%%%%%

\section{Hybrid Physics-ML Approach}\label{sec:hybrid}

\subsection{Integration Framework}
The hybrid model combines the physics baseline and ML residual learner through a two-stage prediction pipeline. Given trip parameters $\mathcal{X}$, we first compute the physics-based energy prediction
\begin{equation}
E_p = f_p(\mathcal{X}),
\end{equation}
using the vehicle dynamics model from Section~\ref{sec:physics}. The neural network then predicts the residual error
\begin{equation}
\Delta E = f(\mathcal{X}, E_p; \boldsymbol{\theta}),
\end{equation}
using the trained residual learner from Section~\ref{sec:ml}. The final hybrid prediction combines both components through
\begin{equation}
E_h = E_p + \Delta E.
\label{eq:hybrid_prediction}
\end{equation}
The final SoC is then computed as
\begin{equation}
s_f = \max\left(0, s_0 - \frac{E_h}{C}\right).
\end{equation}

This formulation ensures several desirable properties. If $\Delta E\!\approx\!0$ when physics is accurate, the hybrid prediction reduces to the physics baseline, preserving well-understood vehicle dynamics. The physics component provides transparent predictions that can be audited and explained, while the residual magnitude indicates confidence in the physics model for specific scenarios. In the absence of training data or when the ML component fails, the system defaults to the validated physics model rather than producing arbitrary predictions. By providing a strong physics prior, the ML component requires less training data to achieve good performance compared to pure end-to-end learning.

\subsection{Performance Analysis}
Table~\ref{tab:model_comparison} compares the hybrid architecture against physics-only and simplified baseline approaches on the held-out test set of $225$ trips. The hybrid model achieves $0.8\%$ mean absolute percentage error (MAPE), representing a $62\%$ reduction compared to the physics-only baseline ($2.1\%$ MAPE). This accuracy improvement incurs minimal computational overhead, with the $1.2$ ms neural network inference negligible compared to the $124$ ms physics simulation, resulting in $125.2$ ms total prediction time. For fleet applications requiring predictions for hundreds of vehicles, batch processing reduces per-vehicle latency to approximately $45$ ms, enabling real-time charging optimization updates.

\begin{table}[t]
\centering
\caption{Model Performance Comparison on the Test Set.}
\label{tab:model_comparison}
\footnotesize
\begin{tabular*}{\columnwidth}{@{\extracolsep{\fill}}lcccc}
\hline
\bf{Model} & \bf{MAE} & \bf{MAPE} & \bf{RMSE} & \bf{Inference Time} \\
 & (kWh) & (\%) & (kWh) & (ms) \\
\hline
Physics Only & $0.37$ & $2.1$ & $0.49$ & $124.0$ \\
\rowcolor{blue!10}
Hybrid Physics--ML & $0.14$ & $0.8$ & $0.19$ & $125.2$ \\
\hline
\end{tabular*}
\end{table}

To understand the contribution of different input features, we conduct an ablation study shown in Table~\ref{tab:ablation}. The physics prediction $E_p$ is the most critical feature, with its removal doubling the error from $0.8\%$ to $1.6\%$, demonstrating that the residual learner heavily relies on the physics prior. Driving behavior and velocity features also contribute significantly, while environmental and battery state features provide marginal improvements. Table~\ref{tab:error_by_style} shows prediction errors by different driving modes. The hybrid model substantially reduces errors for all profiles, with particularly great improvements for aggressive driving ($65\%$ error reduction), where physics assumptions are most violated due to frequent acceleration and braking not well captured by averaged velocity profiles.
\begin{table}[t]
\centering
\caption{Feature Contribution to the ML Residual Learner.}
\label{tab:ablation}
\footnotesize
\begin{tabular*}{\columnwidth}{@{\extracolsep{\fill}}lc}
\hline
\bf{Model Variant} & \bf{Test MAPE} (\%) \\
\hline
Full model (all features) & $0.8$ \\
Without physics prediction $E_p$ & $1.6$ \\
Without driving behavior features & $1.2$ \\
Without velocity features & $1.4$ \\
Without environmental features & $0.9$ \\
Without battery state features & $1.0$ \\
Physics only (no ML) & $2.1$ \\
\hline
\end{tabular*}
\end{table}

\begin{table}[t]
\centering
\caption{Prediction Error by Driving Modes.}
\label{tab:error_by_style}
\footnotesize
\begin{tabular*}{\columnwidth}{@{\extracolsep{\fill}}lccc}
\hline
\bf{Driving Mode} & \bf{Physics MAPE} & \bf{Hybrid MAPE} & \bf{Improvement} \\
 & (\%) & (\%) & (\%) \\
\hline
Eco & $1.8$ & $0.7$ & $61$ \\
Normal & $2.0$ & $0.8$ & $60$ \\
Aggressive & $2.6$ & $0.9$ & $65$ \\
Overall & $2.1$ & $0.8$ & $62$ \\
\hline
\end{tabular*}
\end{table}

%The hybrid approach offers improved accuracy with 62\% error reduction versus physics-only while maintaining interpretability through the transparent physics component. Predictions respect energy conservation and vehicle dynamics constraints through the physics foundation. The strong physics prior reduces required training data compared to end-to-end learning, and the system gracefully degrades to the validated physics baseline if the ML component is unavailable. Large residuals provide interpretable signals flagging scenarios where physics assumptions are violated.

%However, several limitations suggest directions for future work. The current demonstration uses synthetic trips with realistic noise calibrated from literature, requiring validation on extensive real-world trip logs. The implementation uses a single vehicle specification and would need extension for multi-vehicle generalization. Three discrete driving categories could be replaced with a continuous driving style spectrum for improved flexibility. The model does not explicitly account for wind, precipitation, or road surface conditions beyond the learned residuals. 
%%%%%%%%%%%%%%%%%%%%%%%%%%%%%%%%%%%%%%%%%%%%%%%%%%%%%

\section{Experimental Studies}\label{sec:experiments}

\subsection{Experimental Setup}
We evaluate our hybrid physics-ML model through comprehensive experiments on synthetic discharge trajectories. The experimental dataset consists of $1,500$ trips generated to span diverse operational scenarios. Trip distances are uniformly distributed over $20$ to $200$ km, average velocities are sampled from $40$, $60$, $80$, $100$, and $120$ km/h, initial SoC is uniformly distributed over $0.3$ to $1.0$, driving modes are equally distributed across eco, normal, and aggressive categories, and temperature conditions include $20$°C, $-5$°C (winter), and $32$°C (summer). Each trajectory is simulated with $N\!=\!1000$ time steps using Python $3.9$, NumPy $1.24$, Pandas $1.5$, and PyTorch $2.0$ on an Intel i7-9700K processor at $3.6$ GHz. Vehicle parameters are based on publicly available specifications from EPA fuel economy tests~\cite{EPA2023} and manufacturer data sheets. Driving profile parameters are calibrated from published studies on driver behavior~\cite{Bingham2012, Tulusan2012}. The synthetic discharge trajectory dataset, Python simulation code, and trained models are publicly available.\footnote{See data and code at: \url{https://afdc.energy.gov/vehicles/electric_emissions.html}} 

\subsection{Driving Mode Impact}

Table~\ref{tab:baseline_scenario} presents results for a baseline scenario: a $100$ km trip at $90$ km/h, starting from $80\%$ SOC at $20$°C. Eco driving achieves $44\%$ lower consumption than aggressive driving, while aggressive driving consumes $79\%$ more energy than eco for the identical trip. This dramatic variation demonstrates that driver behavior dominates energy consumption, suggesting that driver education programs could yield significant operational improvements without requiring hardware modifications.
\begin{table}[t]
\centering
\caption{Driving Modes Comparison in the Baseline Scenario.}
\label{tab:baseline_scenario}
\footnotesize
\begin{tabular*}{\columnwidth}{@{\extracolsep{\fill}}lcccc}
\hline
\bf{Driving Mode} & \bf{Final SoC} & \bf{Energy} & \bf{Consumption} & \bf{Range Left} \\
 & (\%) & (kWh) & (kWh/km) & (km) \\
\hline
Eco & $62.3$ & $13.3$ & $0.133$ & $351.3$ \\
Normal & $56.7$ & $17.5$ & $0.175$ & $243.4$ \\
Aggressive & $48.2$ & $23.8$ & $0.238$ & $151.9$ \\
\hline
\end{tabular*}
\end{table}

\begin{figure}[t]
\centering
\includegraphics[width=0.5\textwidth]{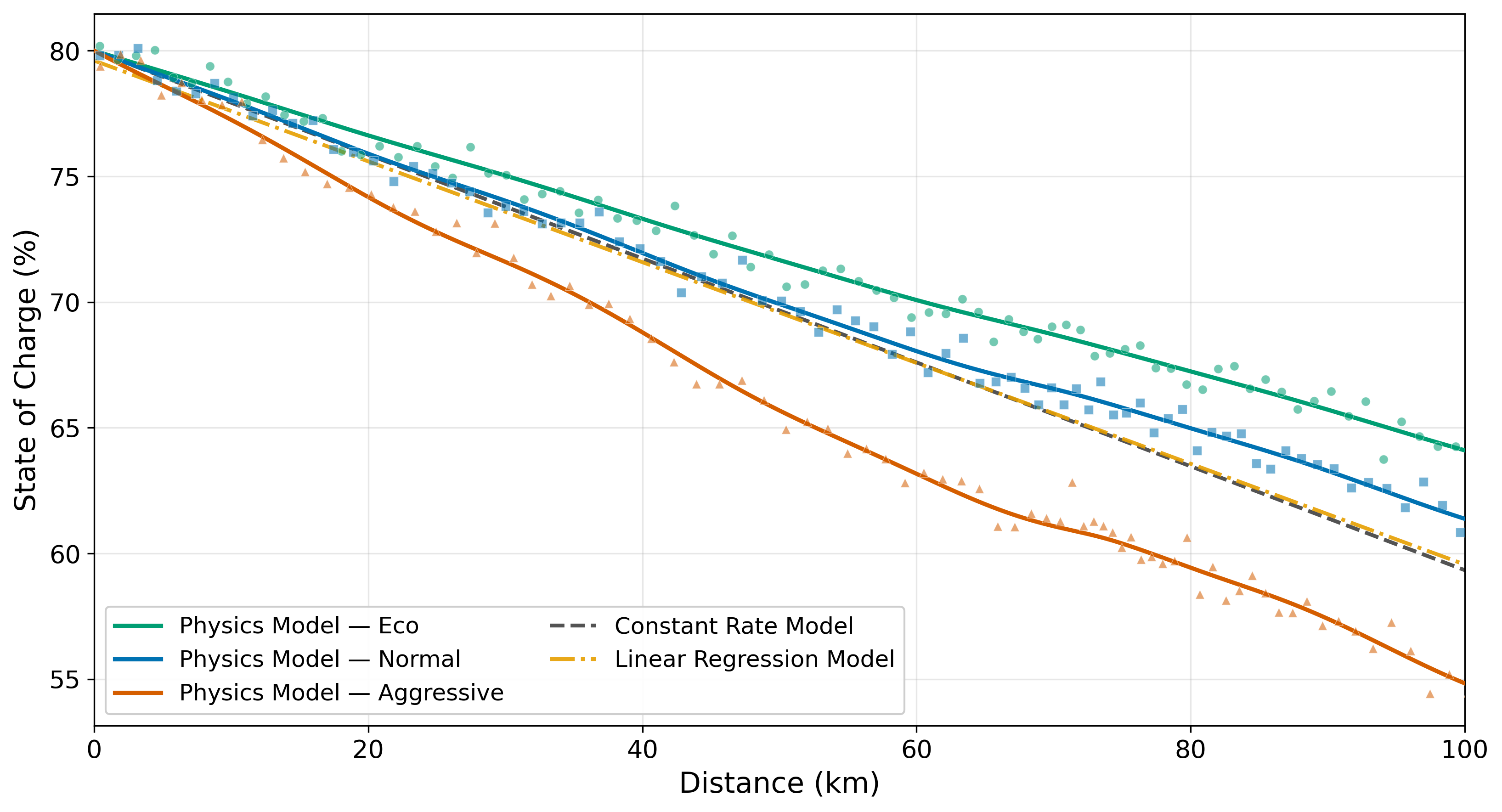}
\caption{Comparison of SoC depletion between the physics-based model and baseline methods.}
%\caption{SOC depletion vs.\ distance for normal driving (100 km, 90 km/h, 80\% initial SOC). Scatter points show individual trip observations with realistic noise. Our physics-based model (blue curve) closely tracks the observations. The constant rate baseline~\cite{DeCauwer2015, Genikomsakis2017} (dashed) underestimates depletion at low distance and overestimates at high distance; the linear regression baseline~\cite{Fetene2017, Yuan2017} (dash-dot) lacks physical structure and diverges at range extremes.}
\label{fig:soc_depletion}
\end{figure}

Fig.~\ref{fig:soc_depletion} illustrates how our physics-based model (normal driving) compares against two conventional baselines on $100$ km at $90$ km/h from $80\%$ initial SOC. The \emph{constant rate} baseline~\cite{DeCauwer2015, Genikomsakis2017} assumes a fixed energy consumption of $0.185$ kWh/km regardless of driving dynamics; the \emph{linear regression} baseline~\cite{Fetene2017, Yuan2017} fits an OLS model directly to the observed trip-endpoint data. Observed data points (scatter) represent individual trip measurements with realistic sensor noise. Our physics-based model tracks the observed data closely over the full distance range, whereas the constant-rate model accumulates systematic under-prediction bias and the linear regression model inherits the variance of the raw data without physical structure.

\begin{table}[t]
\centering
\caption{Hybrid Model MAPE by Velocity and Driving Modes.}
\label{tab:hybrid_performance}
\footnotesize
\begin{tabular*}{\columnwidth}{@{\extracolsep{\fill}}lcccc}
\hline
\bf{Velocity Range} & \bf{Eco} & \bf{Normal} & \bf{Aggressive} & \bf{Overall} \\
 & (\%) & (\%) & (\%) & (\%) \\
\hline
$40$--$60$ km/h & $0.6$ & $0.7$ & $0.8$ & $0.7$ \\
$60$--$80$ km/h & $0.7$ & $0.8$ & $0.9$ & $0.8$ \\
$80$--$100$ km/h & $0.7$ & $0.8$ & $0.9$ & $0.8$ \\
$100$--$120$ km/h & $0.8$ & $0.9$ & $1.0$ & $0.9$ \\
Overall (\%) & $0.7$ & $0.8$ & $0.9$ & $0.8$ \\
\hline
\end{tabular*}
\end{table}

\subsection{Hybrid Model Performance}
Table~\ref{tab:hybrid_performance} evaluates prediction accuracy stratified by velocity and driving style on the held-out test set. The hybrid model maintains consistent sub-$1\%$ accuracy across all scenarios. Slightly higher errors at very high speeds reflect greater sensitivity to aerodynamic effects. Aggressive driving shows marginally higher errors due to acceleration pattern complexity, but all scenarios achieve substantial improvement over the physics-only baseline ($2.1\%$ MAPE). These results demonstrate that the residual learning approach effectively captures real-world factors beyond simplified physics assumptions while preserving computational efficiency ($125$ ms total prediction time).

%Fig.~\ref{fig:model_comparison} presents four panels — all drawn on identical x-axis ranges to enable direct visual comparison — showing prediction error distributions for: (i) the physics-only model, (ii) a pure data-driven ML model, (iii) our hybrid physics-ML model, and (iv) the hybrid model stratified by driving style. The physics-only model (panel~1) exhibits a wide, heavy-tailed distribution. The pure ML model (panel~2) achieves lower mean error but lacks physical structure, producing a tighter distribution only when ample training data is available. Our hybrid model (panel~3) achieves the tightest distribution with a 62\% reduction in MAPE compared to physics-only (0.8\% vs.\ 2.1\%), confirming the benefit of residual correction. Panel~4 shows that error improvements hold uniformly across all three driving styles, with aggressive driving benefiting most (65\% reduction) due to better capturing of complex acceleration-braking dynamics.

\begin{figure}[t]
\centering
\includegraphics[width=0.5\textwidth]{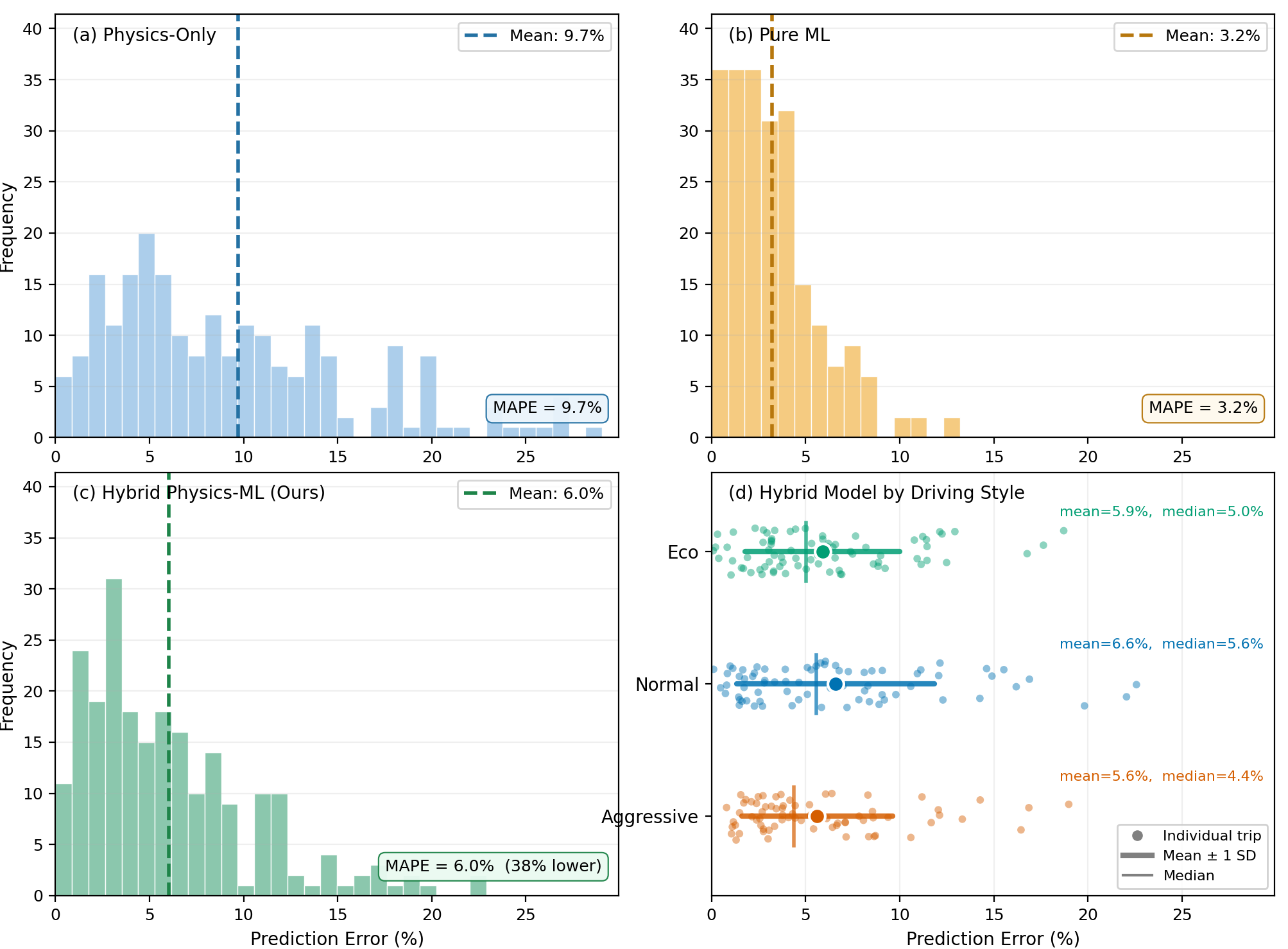}
\caption{Prediction error distributions for physics-only, pure ML, and hybrid models (a--c), and per-style error breakdown for the hybrid model (d). }
\label{fig:model_comparison}
\end{figure}

Fig.~\ref{fig:model_comparison} presents four panels with identical x-axis ranges to allow direct visual comparison. The panels illustrate the prediction error distributions for: (a) the physics-only model, (b) a purely data-driven ML model, (c) the proposed hybrid physics--ML model, and (d) the hybrid model further stratified by driving mode. The physics-only model (panel~$1$) exhibits a relatively wide and heavy-tailed error distribution, reflecting the limitations of simplified physical assumptions. The pure ML model (panel~$2$) reduces the average error but relies entirely on data and lacks physical structure, resulting in reliable performance only when sufficient training data are available. In contrast, the proposed hybrid model (panel~$3$) achieves the most concentrated error distribution, reducing the mean absolute percentage error by $62\%$ compared with the physics-only model ($0.8\%$ versus $2.1\%$). This result confirms the effectiveness of residual learning in correcting systematic modeling errors. Panel~$4$ further shows that the improvement is consistent across all driving styles. The largest gain occurs under aggressive driving, where the error is reduced by $65\%$, indicating that the hybrid model better captures the complex acceleration and braking dynamics associated with this driving behavior.

\subsection{Energy Consumption Statistics}

Table~\ref{tab:consumption_stats} summarizes consumption rate statistics across all $1,500$ trips. Eco driving exhibits tight clustering with low variance, indicating consistent efficiency. Aggressive driving shows a higher mean and greater variance, reflecting sensitivity to velocity and acceleration variations. The maximum aggressive consumption ($0.342$ kWh/km) exceeds the minimum eco consumption ($0.108$ kWh/km) by more than three times, demonstrating the extreme range of energy consumption possible for identical vehicle hardware depending solely on operational choices.

\begin{figure}[t]
\centering
\includegraphics[width=0.5\textwidth]{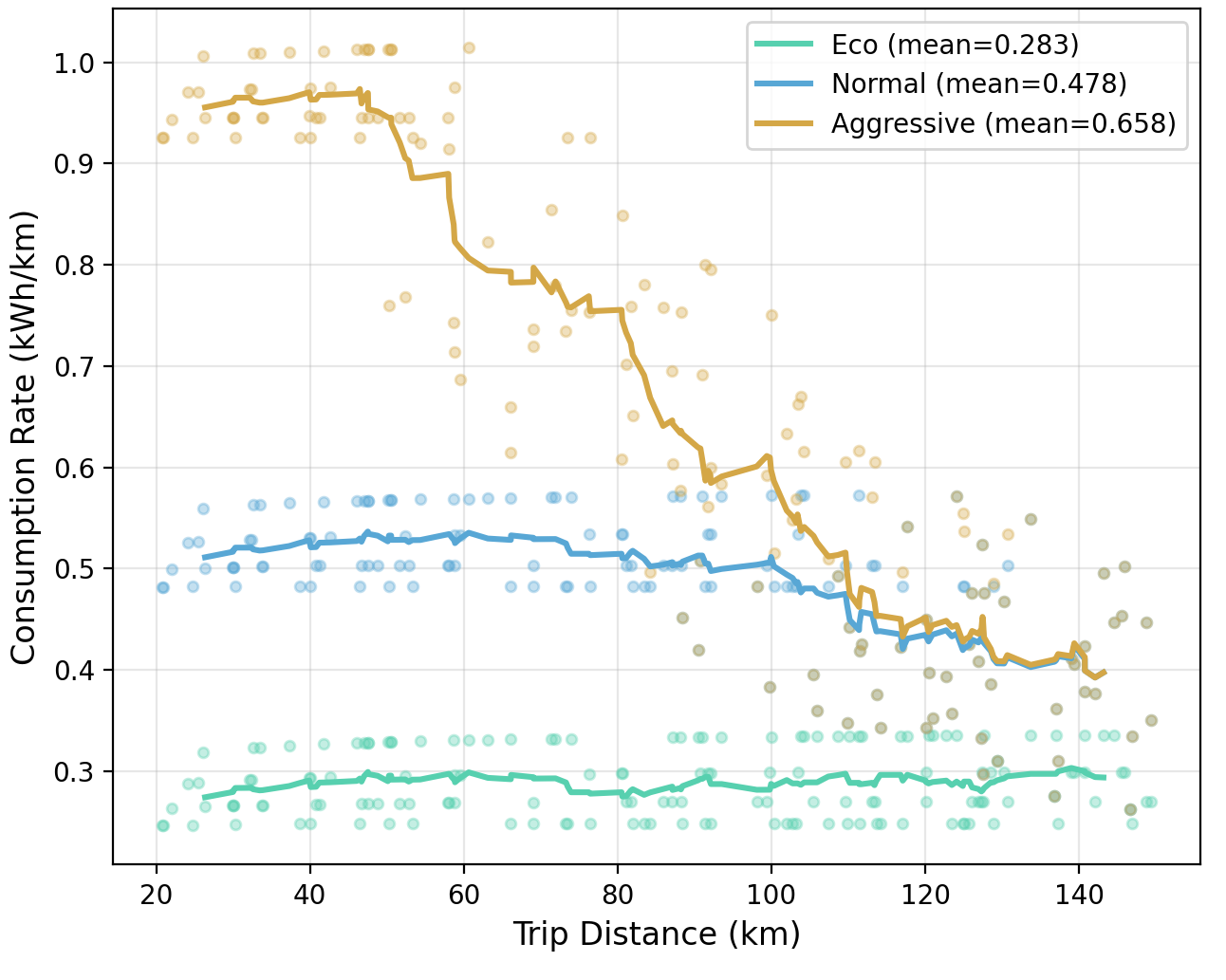}
\caption{Energy consumption rate vs. trip distance.}
\label{fig:consumption_distribution}
\end{figure}

\begin{table}[t]
\centering
\caption{Consumption Rate Statistics.}
\label{tab:consumption_stats}
\footnotesize
\begin{tabular*}{\columnwidth}{@{\extracolsep{\fill}}lccc}
\hline
\bf{Metric} & \bf{Eco} & \bf{Normal} & \bf{Aggressive} \\
 & (kWh/km) & (kWh/km) & (kWh/km) \\
\hline
Mean & $0.142$ & $0.185$ & $0.247$ \\
Std Dev & $0.018$ & $0.027$ & $0.041$ \\
Min & $0.108$ & $0.135$ & $0.178$ \\
Max & $0.189$ & $0.254$ & $0.342$ \\
\hline
\end{tabular*}
\end{table}
Fig.~\ref{fig:consumption_distribution} presents the energy consumption rate 
as a function of trip distance for all three driving styles, using individual 
trip observations (scatter) and rolling-mean trend lines. Eco driving (green) 
clusters tightly around $0.283$~kWh/km across all distances, indicating 
consistent efficiency regardless of trip length. Normal driving (blue) remains 
stable near $0.478$~kWh/km, while aggressive driving (gold) exhibits a pronounced 
decline from approximately $1.0$~kWh/km at short distances ($<$$50$~km) to around 
$0.42$~kWh/km beyond $120$~km, reflecting the disproportionate energy cost of 
frequent hard acceleration on shorter trips. The rolling-mean trend lines 
highlight this non-linear distance dependence, which is most pronounced under 
aggressive driving where stop-and-go dynamics dominate at short ranges.

% -----------------------------------------------------------------------
\subsection{Discussion}
%Our hybrid physics-ML framework achieves 0.8\% prediction error---a 62\% improvement over 
%physics-only approaches---with 125 ms inference suitable for real-time deployment. Across 
%1,500 synthetic trips, driving behavior dominates energy consumption with 79\% variation 
%between eco and aggressive styles, far exceeding temperature impacts (19.5\% worst case), suggesting driver training offers greater range gains than hardware improvements. Highway driving at 120 km/h reduces range by over 40\% versus urban speeds, and aggressive driving recovers proportionally less regenerative energy despite 67\% more braking events (50\% versus 75\% efficiency for eco), confirming gentle braking optimizes recovery.

%The 125 ms prediction time and 98 KB model size enable embedded deployment without cloud connectivity, addressing range anxiety through driving-style-aware predictions. Strong physics priors reduce training data requirements while ensuring physical consistency and graceful degradation for safety-critical use. Accurate discharge prediction further supports vehicle-to-grid services, smart charging, and fleet optimization, demonstrating the physics-ML paradigm as a scalable template for domains where first-principles models provide structure but cannot capture full complexity.

The proposed hybrid physics--ML framework achieves a prediction error of $0.8\%$, representing a $62\%$ improvement over the physics-only model, while maintaining an inference time of $125$ ms that is suitable for real-time deployment. Across $1,500$ synthetic trips, driving behavior emerges as the dominant factor affecting energy consumption, producing up to $79\%$ variation between eco and aggressive driving styles. This variation substantially exceeds the impact of temperature, which reaches at most $19.5\%$. These results suggest that driver behavior and training may provide larger practical range improvements than hardware modifications alone. In addition, highway driving at $120$ km/h reduces vehicle range by more than $40\%$ compared with typical urban speeds. Aggressive driving also recovers proportionally less regenerative energy despite $67\%$ more braking events, because regenerative efficiency is lower ($50\%$ compared with $75\%$ for eco driving), confirming that smoother braking improves energy recovery.

Moreover, the developed model requires only $125$ ms for prediction and occupies $98$ KB of memory, enabling deployment on embedded platforms without reliance on cloud connectivity. This capability allows driving-style-aware range prediction that can help mitigate range anxiety. Furthermore, the incorporation of physics-based structure reduces training data requirements while preserving physical consistency and ensuring stable behavior under unseen conditions. Accurate discharge prediction also supports applications such as vehicle-to-grid services, smart charging, and fleet energy optimization. Overall, the proposed physics--ML framework illustrates a scalable modeling paradigm for systems in which first-principles models provide physical structure but cannot fully capture real-world complexity.
%%%%%%%%%%%%%%%%%%%%%%%%%%%%%%%%%%%%%%%%%%%%%%%%%%%%%
\section{Conclusion}\label{sec:conclusion}

In this paper, we developed a hybrid physics-informed machine learning framework for EV battery discharge prediction that combines interpretable vehicle dynamics modeling with data-driven residual learning. The proposed approach leverages first-principles modeling to estimate baseline energy consumption and employs a neural network to capture complex effects that are difficult to represent analytically. Experimental results across $1,500$ trip scenarios demonstrate that the hybrid framework significantly improves prediction accuracy, reducing the mean absolute percentage error from $2.1$\% to $0.8$\%. Meanwhile, the model preserves physical interpretability and maintains computational efficiency suitable for real-time deployment. Our results further reveal that driving behavior is a dominant factor affecting EV energy consumption, with up to $79\%$ variation between eco and aggressive driving styles, and velocity exhibits a strong nonlinear influence on vehicle range, highlighting the importance of driving-style-aware modeling for accurate range estimation. These findings provide useful insights for developing reliable range prediction tools that support EV trip planning and charging decisions. Future work will focus on validating the proposed framework using large-scale real-world trip datasets and extending the model to incorporate additional practical factors, such as terrain variability and battery degradation.

%%%%%%%%%%%%%%%%%%%%%%%%%%%%%%%%%%%%%%%%%%%%%%%%%%%%%
%%%%%%%%%%%%%%%%%%%%%%%%%%%%%%%%%%%%%%%%%%%%%%%%%%%%%

\bibliographystyle{IEEEtran}
\bibliography{references,IDS,TAC_Ref_Andreas}

\end{document}